# Very High Field Two-Band Superconductivity in LaFeAsO$_{0.89}$F$_{0.11}$

F. Hunte, J. Jaroszynski, A. Gurevich, D. C Larbalestier

*National High Magnetic Field Laboratory, Florida State University, Tallahassee, Florida 32310, USA.*

R. Jin, A.S. Sefat, M.A. McGuire, B.C. Sales, D.K. Christen, D. Mandrus

*Oak Ridge National Laboratory, P.O. Box 2008, Oak Ridge, TN, USA.*

**The ongoing search for new superconductors has recently yielded a new family of oxypnictides composed of alternating La$_2$O$_{2-x}$F$_x$ and Fe$_2$As$_2$ layers [1-4] with transition temperatures $T_c$ of 25-28 K, which can be raised to 40-43 K by replacing La by Ce [5] or Sm [6] or to 52 K by replacing La with Nd and Pr [7, 8]. Recent experiments and band structure calculations have suggested an unconventional multiband superconductivity in the layers of paramagnetic Fe ions, which would normally destroy superconductivity in the traditional mechanism of the s-wave Cooper pairing. Here we report very high-field resistance measurements up to 45T, which show a remarkable enhancement of the upper critical fields $B_{c2}$ at low temperatures, as compared to those expected from the already high slopes of $dB_{c2}/dT \approx$ 2T/K near $T_c$. The deduced $B_{c2}(0) \approx$ 63-65 T exceeds the paramagnetic limit, consistent with strong coupling and important two-band effects in LaFeAsO$_{0.89}$F$_{0.11}$. We argue that oxypnictides are emerging as a new class of high-field superconductors surpassing the $B_{c2}$ of Nb$_3$Sn, MgB$_2$, and the Chevrel phases and perhaps approaching the 100T field benchmark of the high-$T_c$ cuprates.**

The recent synthesis of the novel superconductor LaFeAsO$_{0.89}$F$_{0.1}$ with the transition temperature of T$_c$ = 26K [1-4] has been quickly followed by reports of even higher T$_c$ = 41K in CeFeAsO$_{0.89}$F$_{0.1}$ [5], T$_c$ = 43K in SmFeAsO$_{0.89}$F$_{0.1}$ [6] and T$_c$ = 52K in NdFeAsO$_{0.89}$F$_{0.1}$ and PrFeAsO$_{0.89}$F$_{0.1}$ [7,8] These discoveries have generated much interest in the mechanisms and manifestations of unconventional superconductivity in the family of doped quaternary layered oxypnictides LOMPn (L = La, Pr, Ce, Sm; M = Mn, Fe, Co, Ni; Pn=P, As) [9, 10], because many features of these materials clearly set them apart from other superconductors. First, ab-initio calculations indicate that superconductivity originates from the d-orbitals of what would normally be expected to be pairbreaking magnetic Fe ions, suggesting that new non-phonon pairing mechanisms are responsible for the high T$_c$ [11, 12]. Second, F-doped LaFeAsO is a semimetal, which exhibits strong ferromagnetic and antiferromagnetic fluctuations and a possible spin density wave instability around 150K in the parent undoped LaFeAsO [5,13-16]. And third, superconductivity emerges on several disconnected pieces of the Fermi surface

[11,12,17,18], thus exhibiting the multi-gap pairing, which has recently attracted so much attention in $MgB_2$ [19].

Given the importance of magnetic correlations in the doped oxypnictides, transport measurements at very high magnetic fields are vital to probe the mechanisms of superconductivity. Indeed, first measurements of the upper critical field $B_{c2}$ at low fields B < 7T have yielded a slope $B_{c2}'(T_c) = dB_{c2}/dT \approx 2T/K$ near $T_c$, for both La and Sm based oxypnictides [2-6]. From the conventional one-band Werthamer-Helfand-Hohenberg (WHH) theory [20] such slopes already imply rather high values $B_{c2}(0) = 0.69T_cB_{c2}' \approx$ 36T for $LaFeAsO_{0.89}F_{0.1}$, $B_{c2}(0) \approx 59.3T$ for $SmFeAsO_{0.89}F_{0.1}$, and $B_{c2}(0) \approx 72$ T for $PrFeAsO_{0.89}F_{0.1}$, all well above $B_{c2}$ of $Nb_3Sn$. However, studies of the high-field superconductivity in $MgB_2$ alloys have shown that the upward curvature of $B_{c2}(T)$ resulting from the multiband effects can significantly increase $B_{c2}(0)$ as compared to the WHH one-band extrapolation (see, e.g., the review [21] and references therein). To address these important issues, we have performed high-field dc transport measurements on $LaFeAsO_{0.89}F_{0.1}$ samples up to 45T. We show that $B_{c2}(T)$ indeed exhibits two-gap behavior similar to that in $MgB_2$, and $B_{c2}(0)$ values exceed the WHH extrapolation by the factor ~ 2. Moreover, the observed $B_{c2}(0)$ also exceeds the BCS paramagnetic limit $B_p[T] = 1.84T_c[K] = 47T$ for $T_c = 26K$.

Polycrystalline $LaFeAsO_{0.89}F_{0.11}$ samples were made by solid state synthesis [4]. A sample ~ 3 x 1 x 0.5 mm was used for our four probe transport measurements in the 45T Hybrid magnet at the NHMFL, supplemented by low field measurements in a 9T superconducting magnet. Our low-field data agreed well with the earlier data taken at ORNL on the same sample [4], indicating its good temporal and atmospheric stability. The 45T Hybrid magnet was swept only from 11.5T to 45T due to the constant 11.5T background of the outsert magnet while lower fields were swept from 0T to 9T in a PPMS with resistivity measured in AC mode using a 5mA excitation current, whereas the high field resistance R(B) was measured by a Keithley nanovoltmeter at 2mA. The temperature range from 30K to 2.5K was explored using a rotator probe with particular emphasis on the behavior of $B_{c2}$ at low temperature (<10 K).

The results of our high-field measurements of the sample resistance R(B) are shown in Figure 1. The broad R(B) transitions are not surprising because the sample consists of anisotropic crystalline grains with different orientations of the c-axis. Given the predicted high resistivity ratio $\Gamma = \rho_c/\rho_{ab} \approx 10\text{-}15$ for this layered compound [11], the local $B_{c2}(\theta) \approx B_{c2}(0)[\cos^2\theta + \Gamma^{-1}\sin^2\theta]^{-1/2}$ should vary strongly, depending on the angle $\theta$ between the c-axis in the grain and the applied field. Thus, we can identify two characteristic fields: the high field onset $B_{max}$ of the superconducting transition, and the zero-resistance, low-field onset $B_{min}$, as illustrated by Figure 1. The in-field transitions are shown in Figure 1a for B perpendicular to the broad 1 mm wide face of the sample and Figure. 1b for B applied along the shortest sample dimension (in both cases transport current was always perpendicular to **B**). A difference in the R(B) curves seen in Figures 1a and 1b indicates some grain texture in the sample, so the zero-resistance field $B_{min}$ in Figure 1b is higher than in Figure 1a. For example, at 4.2K, we have $B_{min} = 25T$ and 36T, respectively, suggesting that the plate-like grains tend to align their ab planes parallel to the broad face of the sample.

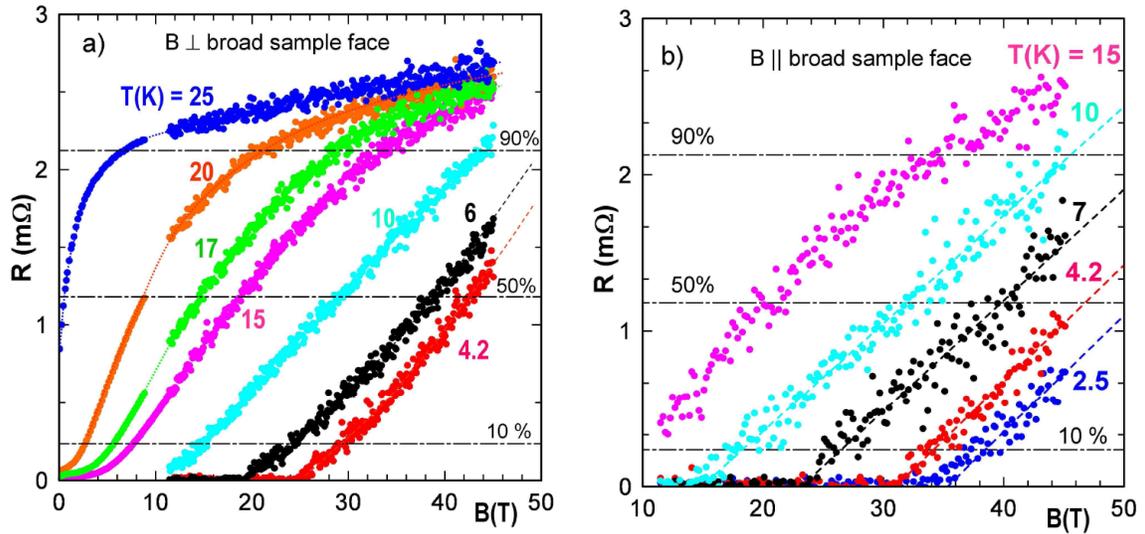

**Figure 1 (a)** The resistance R(B) for different temperatures taken in swept fields in the 45T hybrid magnet at a measuring current of 2 mA and from 0-9T in a superconducting magnet at a measuring current of 5 mA for B applied perpendicular to the broad face of the sample. **(b)** R(B) data taken in the hybrid magnet at different temperatures for B parallel to the broad sample face, which we believe has a higher fraction of ab-oriented grains.

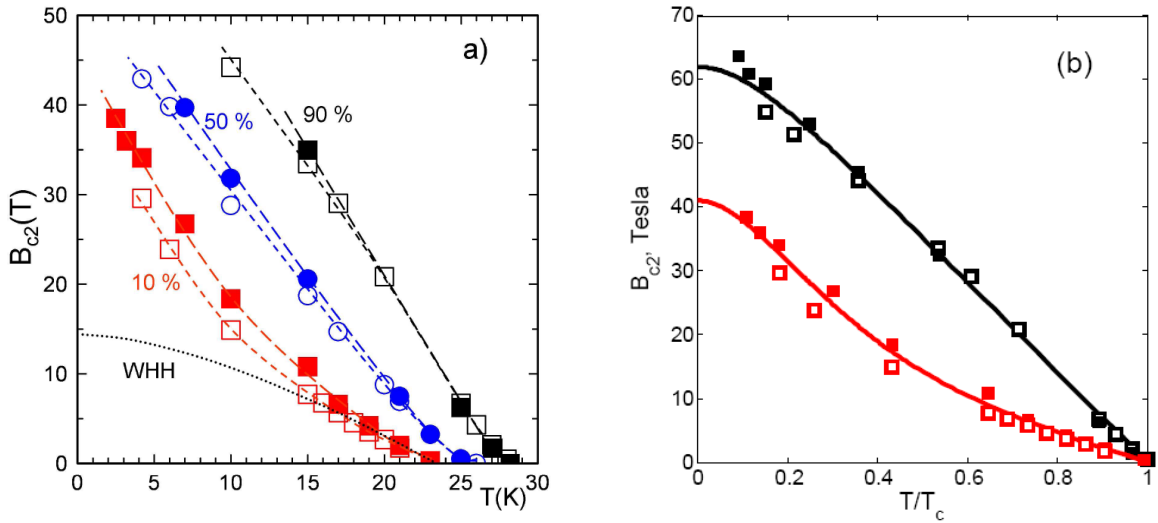

**Figure 2.** **(a)** The measured fields $B_{max}(T)$ and $B_{min}(T)$ (squares) and the midpoint transition fields (circles). The full and open symbols correspond to the parallel and perpendicular field orientations. The dashed line shows the WHH curve defined by the slope of $B_{min}(T)$ at $T_c$. **(b)** $B_{max}(T)$ (black) and $B_{min}(T)$ (red) plotted as functions of the reduced temperature $T/T_c$. The data points above 45T were extracted by linear extrapolation of R(B) at B < 45T to R(B) = $0.9R_n(T_c)$ as shown by dashed lines in Figure 1. The lines correspond to $B_{c2}(T)$ calculated from the two-gap theory for the parameters described in the text.

Shown in Figure 2a are the temperature dependencies of the fields $B_{min}(T)$, $B_{mid}(T)$ and $B_{max}(T)$ extracted from the measured curves R(B) evaluated at 90%, 50%, and 10% of the normal state resistance $R_n(T_c)$, respectively. The interpretation of $B_{min}$ and $B_{max}$ in polycrystals is often complicated by strong vortex pinning and the particulars of the grain misorientation distribution. However in our case, analysis is greatly simplified because: 1. The measured magnetization curves are nearly reversible, suggesting weak pinning. 2. Crystalline anisotropy of the layered $LaFeAsO_{0.89}F_{0.1}$ compound results in a very broad distribution of the local $B_{c2}(\theta)$ values of different grains. Thus, $B_{max}(T)$ can be associated with the larger in-plane upper critical field $B_{c2}^{\parallel}(T) \approx B_{c2}^{\perp}(T)\Gamma^{1/2}$ because grains with their ab planes oriented along the applied field first become superconducting upon cooling. In turn, the zero-resistance field $B_{min}(T)$ can be interpreted as the field, below which those superconducting grains with $B_{c2}(\theta) > B$ form a percolative path at the low measuring current density $\approx 0.2$ A/cm$^2$. For $\Gamma \gg 1$ and negligible thermal activation of vortices, the effective medium theory then gives $B_{min}(T) \approx p_c B_{c2}^{\perp}(T)$, where $p_c = 1 - x_c$, and $x_c \sim 1/3$ is the 3D percolation threshold. Therefore, we conclude that $B_{max}(T)$ and $B_{min}(T)$ reflect the temperature dependencies of $B_{c2}^{\parallel}(T)$ and $B_{c2}^{\perp}(T)$, respectively.

As is evident from Figure 2, $B_{c2}^{\perp}(T)$ exhibits a significant upward curvature, which is much less pronounced for $B_{c2}^{\parallel}(T)$. Because this behavior is strikingly similar to that observed on MgB$_2$ dirty films [21], we can conclude, in agreement with the ab-initio calculations [11, 12], that superconductivity in $LaFeAsO_{0.89}F_{0.1}$ indeed results from two bands: a nearly 2D electron band with high in-plane diffusivity $D_1$ and a more isotropic heavy hole band with smaller diffusivity $D_2$. The upward curvature of $B_{c2}(T)$ is then controlled by the ratio $\eta = D_2/D_1$: for $\eta \ll 1$, the upward curvature is pronounced, while for $\eta \approx 1$, $B_{c2}(T)$ exhibits a more traditional WHH-like behavior [20]. For field along the ab plane, the parameter $\eta$ should be replaced with $\eta = D_2/[D_1^{(ab)}D_1^{(c)}]^{1/2}$, allowing strong anisotropy $D_1^{(ab)} \gg D_1^{(c)}$ to significantly increase $\eta$ for B||ab as compared to B||c [21]. In this case the upward curvature of $B_{c2}^{\parallel}(T)$ does become less pronounced than for $B_{c2}^{\perp}(T)$, in agreement with the results shown in Figure 2.

To check the self-consistency of our interpretation, we fit $B_{c2}^{\perp}(T)$ in Figure 2, using the two-gap theory outlined in the supplemental material. We took $\eta = 0.08$ and the interband BCS coupling constants $\lambda_{12} = \lambda_{21} = 0.5$, but the results turn out not to be particularly sensitive to the choice of either $\lambda_{12}$ and $\lambda_{21}$ or the intraband coupling constants $\lambda_{11}$ and $\lambda_{22}$. For example, Figure 2 shows a rather good fit of the theory to the data for the case of the strong interband repulsion $\lambda_{12}\lambda_{21} \gg \lambda_{11}\lambda_{22}$ suggested in Ref. 12. However, using Eqs. (1)-(3) of the supplemental material, we can also show that the fit remains nearly as good, even if we assume that intraband pairing is significant $\lambda_{11}\lambda_{22} > \lambda_{12}\lambda_{21}$ with all coupling constants being of the same order of magnitude. Thus, our experimental data do not enable us to distinguish between different pairing scenarios suggested in the literature, yet they do indicate significant difference in the effective masses in the electron and hole bands. For B||ab, we rescaled the parameter $\eta \rightarrow \eta\Gamma^{1/2} \approx 0.31$, taking the estimate $\Gamma = \rho_c/\rho_{ab} \approx 15$ suggested in Ref. [11], which describes $B_{c2}^{\parallel}(T)$ well, as is evident from Figure 2. Therefore the two-gap scenario is qualitatively consistent with our experimental data.

In conclusion, the newly discovered $LaFeAsO_{0.89}F_{0.11}$ shows exceptionally high $B_{c2}$. Even more remarkable is that this obviously non-optimized material has been quickly synthesized in bulk form showing zero resistance above 35T at 2.5K! Moreover, given the high values of the slope $B_{c2}' \approx$ 2-3 T/K which have also been observed on the Sm-based oxypnictide with $T_c$ = 43 K [6] and Pr or Nd based compounds with $T_c$ = 52K, we could suggest that the values $B_{c2}^{\parallel}(T)$ and $B_{c2}^{\perp}(T)$ would likely increase by the factor 1.5-2 as compared to LaFeAsO$_{0.89}$F$_{0.11}$. Thus, based on the recent experimental data of several groups, as well as the results reported in this work, we can conclude that doped oxypnictides could become a new family of high-field superconductors for which extensive high-field pulse measurements will be required to fully reveal the novel physics of competing superconducting and magnetic orders. This will be of great scientific interest and may also have great importance for high-field applications.

**Acknowledgements.** Work at the NHMFL was supported by IHRP 227000-520-003597-5063 under NSF Cooperative Agreement DMR-0084173, by the State of Florida, by the NSF Focused Research Group on Magnesium Diboride (FRG) DMR-0514592 and by AFOSR under grant FA9550-06-1-0474. Work at ORNL was supported by the Division of Materials Science and Engineering, Office of Basic Energy Sciences under Contract No. DE-AC05-00OR22725. We are very grateful for discussions with G. Boebinger, E. Hellstrom, P. Lee, J. Jiang and C. Tarantini at the NHMFL.

## Supplemental material: the two gap theory.

To analyze the experimental data, we use the theory of two-gap s-wave pairing, which has been recently applied to describe the behavior of $MgB_2$ (see, e.g., a review Ref. [21] and references therein). For the sake of simplicity, we neglect here the paramagnetic effects and interband impurity scattering. In which case $T_c = 1.14\omega_0\exp[(\lambda_0 - \lambda_+)/2w]$, where $\omega_0$ is the cutoff frequency of the exchange boson, $\lambda_\pm = \lambda_{11} \pm \lambda_{22}$, $\lambda_0 = (\lambda_-^2 + 4\lambda_{12}\lambda_{21})^{1/2}$, $w = \lambda_{11}\lambda_{22} - \lambda_{12}\lambda_{21}$, $\lambda_{mn}$ is the 2×2 matrix of the BCS coupling constants where the diagonal elements $\lambda_{11}$ and $\lambda_{22}$ quantify the pairing strength in the bands 1 and 2 and the off-diagonal components $\lambda_{12}$ and $\lambda_{21}$ describe the interband coupling. For $w < 0$, superconductivity can result from strong interband repulsion giving rise to the interband $\pi$ shift between the order parameters on the disconnected parts of the Fermi surface [12]. $H_{c2}$ is described by the following parametric equation [21]

$$\ln\frac{T_c}{T} = \frac{1}{2}\left[U(s) + U(\eta s) + \frac{\lambda_0}{w}\right] - \left\{\frac{1}{4}\left[U(s) - U(\eta s) - \frac{\lambda_-}{w}\right]^2 + \frac{\lambda_{12}\lambda_{21}}{w^2}\right\}^{1/2}, \quad (1)$$

$$H_{c2} = 2\phi_0 Ts/D_1, \qquad \eta = D_2/D_1, \qquad (2)$$

$$U(s) = \psi(s + 1/2) - \psi(1/2), \qquad (3)$$

where the change of the parameter s from 0 to ∞ corresponds to the change of T from $T_c$ to 0, $\psi(x)$ is the di-gamma function, $D_1$ and $D_2$ are the electron (hole) diffusivities in the bands 1 and 2, and $\phi_0$ is the magnetic flux quantum. Equations (1-3), used to fit the data shown in Figure 2b enable us to take into account the effects of very different mobility and anisotropy in the heavy 3D hole band and 2D electron band in LaFeOF compounds. For B||ab, the anisotropic diffusivity $D_1$ should be replaced with $D_1 \to [D_1^{(ab)}D_1^{(c)}]^{1/2}$.